\documentclass[a4paper]{llncs}
\usepackage{url}
\usepackage{color}
\usepackage{graphicx}
\usepackage{amsmath}
\usepackage{float}
\usepackage{url}
\usepackage{hyperref}
\usepackage{caption}
\usepackage{subcaption}
\usepackage{listings}
\usepackage{amsmath}
\usepackage{amssymb}
\usepackage{todonotes}
\usepackage[title]{appendix}

\def\BibTeX{{\rm B\kern-.05em{\sc i\kern-.025em b}\kern-.08em
    T\kern-.1667em\lower.7ex\hbox{E}\kern-.125emX}}

\begin{document}

%\title{Continuous Formal Verification\\{\large Towards a Scalable Approach for Model Checking Complex Software Systems}}
\title{Boost the Impact of Continuous\\Formal Verification in Industry}
%\title{Towards a Scalable Approach for Model Checking Complex Software Systems at Scale}

\author{Felipe R. Monteiro$^{1}$ \and
        Mikhail R. Gadelha$^2$ \and
        Lucas C. Cordeiro$^{1, 3}$
        }
\authorrunning{M. R. Gadelha, F. R. Monteiro, and L. C. Cordeiro}
\institute{
  $^1$Federal University of Amazonas, Brazil, \url{felipemonteiro@ufam.edu.br}\\
  $^2$SIDIA Instituto de Ci\^encia e Tecnologia, Brazil, \url{m.gadelha@samsung.com}\\
  $^3$University of Manchester, UK, \url{lucas.cordeiro@manchester.ac.uk}
  %\url{esbmc@googlegroups.com}
}
\maketitle

\begin{abstract}
Software model checking has experienced significant progress in the last two decades, however,
one of its major bottlenecks for practical applications remains its scalability and adaptability.
Here, we describe an approach to integrate software model checking techniques
into the DevOps culture by exploiting practices such as continuous integration and
regression tests. In particular, our proposed approach looks at the modifications
to the software system since its last verification, and submits them to a
continuous formal verification process, guided by a set of regression test cases.
Our vision is to focus on the developer in order to integrate formal verification techniques
into the developer workflow by using their main software development methodologies and tools.

%Our preliminary evaluation, consisting of a case study from the telecommunications domain,
%shows that the proposed approach can improve error-detection
%and reduce overall verification time when checking software vulnerabilities.
\end{abstract}

\begin{keywords}
Formal Software Verification, Model Checking, DevOps.
\end{keywords}

%-------------------------------------------
\section{Motivation}
\label{sec:motivation}
%-------------------------------------------

%Currently, our software engineering community faces a pressing problem to ensure security of services that are provided by the connection of embedded systems to the Internet. In particular, those systems represent the building block of the Internet of Things (IoT) and it is estimated that billions of embedded devices will be connected to the internet in a near future, which raises deeper concerns about their security~\cite{Kopetz2011}. The complexity of software in IoT devices has increased significantly over the last years so that software verification plays an important role in ensuring the overall product quality. In this context, bounded model checking (BMC)~\cite{Biere03} has been successfully applied to discover subtle errors, but for larger applications, it often suffers from the state space explosion problem~\cite{esbmc2018}.

%problem
Currently, the formal verification community faces a pressing problem to ensure security and reliability of large codebases, which have a significant impact in millions of users~\cite{OHearn:2018}. Even minor defects can lead to huge impacts for companies and costumers~\cite{SadowskiAEMJ18}; for instance, in September 2018, attackers exploited three Facebook vulnerabilities and stole access tokens from as many as 50 million users, in order to take over their accounts~\cite{facebook50m}. In this particular context, software verification plays an important role in ensuring the overall product reliability. Even though formal techniques have been dramatically evolved over the past $15$ years, the main challenges in the formal methods community remain {\bf scalability and adoptability}~\cite{Clarke2018}. {\it So how can we scale formal verification techniques for real-world software systems? How can we increase adoption of formal verification techniques by software engineers in industry?}

%vision
In order to tackle both aforementioned questions, {\bf our vision} is to integrate formal verification techniques into the workflow of the main software development methodologies and tools. Our work is inspired by recent insights described by Sadowski et al.~\cite{SadowskiAEMJ18} who describe a set of lessons from building static analysis tools at Google. We believe that formal methods can be effective in improving software quality assurance of a large number of organisations around the globe. In particular, our approach aims to provide a solution that applies formal verification in a way that is both {\it (i)} low-effort {\it e.g.}, fits into existing processes, and {\it (ii)} scalable to the large software systems used in industry. Here our focus is on software model checking techniques combined with DevOps culture, particularly, continuous integration (CI). On one hand, we have software model checking~\cite{Clarke2018}, which has been successfully applied to discover subtle errors but, for larger applications, often suffers from the state-space explosion problem~\cite{esbmc2018}. On the other hand, we have continuous integration, which has been widely adopted by the software development community, but relies on a test suite that typically does not cover significant parts of the state-space~\cite{Zhao2017}.

%approach
We propose a continuous formal verification (CFV) approach, which aims to automatically detect design errors and integration problems as quickly as possible. First, we concentrate the verification effort to code changes rather than the entire system, thus, we only re-verify the code changes that could potentially break the properties of a system; this verification process should run fast ({\it e.g.}, in less than 5 minutes) in order to provide quick (and useful) feedback for developers. Second, we select the regression tests related to each code change ({\it e.g.}, an updated function), generalize these tests, and formally verify the code changes using software model checking. Lastly, we gather all the information from this analysis and report it back to the analytic and development team, who will carry out this process continuously; this step is crucial according to Sadowski et al. since careful developer workflow integration is key for any static analysis tool adoption by engineers~\cite{SadowskiAEMJ18}.

%results
%As a proof of concept, we conduct a case study on a commercial telecommunication product, which demonstrates the feasibility of our approach to check for vulnerabilities ({\it cf.} Section~\ref{sec:preliminary-results}).

%contributions
%Our {\bf main contributions} are twofold, propose an approach that feasible integrates model checking into DevOps practices, in order to disseminate formal techniques to software engineers; and reduce the impact of the state-space explosion problem in development practices such as CI.
Our {\bf main contributions} are twofold. Firstly, we propose a feasible integration of software model checking into DevOps practices, thus making formal verification techniques accessible to software engineers. Here, our approach will focus on the developer and their feedback; the goal is to increase the adoption of our approach in real-world software projects by integrating our verification tool into the developer workflow. Secondly, we propose to reduce the impact of state-space explosion in development practices using existing regression tests in the verification process, which will provide quick and useful feedback for developers so that they can easily locate and fix bugs.
\section{Continuous Formal Verification}
\label{sec:continuous}
%----------------------------------------------------------------

The essence of this approach relies on the principle of compositional analysis~\cite{OHearn:2018}. Practically, we are inspired by CI practice, a well-known concept in Extreme Programming, proposed by Martin Fowler~\cite{Zhao2017}. CI is particularly relevant when coupled with tools to automatically build and test a project's code base. Since the builds are generated after every incoming code changes ({\it i.e.}, commits and pull requests), problems can be detected much earlier. %it is easier to detect problems much faster.
We can take advantage of such modularity to apply formal techniques in a continuous environment by model checking a software component only after it is changed, {\it i.e.}, we place our approach at diff-time. We use the same information ({\it i.e.}, development history and regression test cases), but in a way to substantially reduce verification complexity and increase coverage in a pull-based development model ({\it e.g.}, GitHub\footnote{More info at https://help.github.com/articles/about-pull-requests/}).

%\begin{figure*}[ht]
%\centering
%\includegraphics[scale=0.35]{figures/approach-overview.pdf}
%\caption{Overview of the continuous formal verification approach.}
%\label{figure:continuous-verification}
%\end{figure*}

The development cycle initiates with the developer submitting changes to the code base through a software configuration management (SCM) system. For each system build, we thus use the information from the SCM system to identify the components that have actually been modified and focus on these. Importantly, we focus on C projects and each function is considered as a component. Equivalence checking~\cite{Strichman09} is then performed to identify which changes have an actual impact on the code base. At this point, the regression test suit (containing unit and functional tests) is of paramount importance, since we select the regression tests correspondent to the non-equivalent changed components. To increase coverage, these regression tests are passed to a generalization process. Finally, we check the use of non-equivalent components through the generalized regression tests and collect the reports ({\it e.g.}, counterexamples) and send it to the analytics team. In order to adopt such an approach, a project must comply with two basic guidelines: {\it (i)} the development process must be based on a continuous integration environment and {\it (ii)} it must include a regression test suite. We are currently building tools that can completely automate our CFV process.

The following sections describe the two steps of the process, highlighting key challenges: identifying relevant code changes, and the model checking of generalized test cases. As an illustrative example, we use the a GitHub project called {\tt vec}\footnote{Available at {\tt https://github.com/rxi/vec}}, an ANSI-C type-safe dynamic array implementation. The repository contains $22$ test cases, which \emph{intend} to cover all possible execution paths related to the functionalities of the type-safe dynamic array. We focus on the function \texttt{vec\_insert\_}, shown in Fig.~\ref{fig:ansi-c-example}, to exemplify our proposed approach.

\begin{figure}[ht]
\centering
\begin{minipage}{.96\textwidth}
\begin{lstlisting}
#define vec_unpack_(v) \
  (char**)&(v)->data, &(v)->length, &(v)->capacity, sizeof(*(v)->data)

#define vec_insert(v, idx, val) \
  ( vec_insert_(vec_unpack_(v), idx) ? -1 : \
    ((v)->data[idx] = (val), 0), (v)->length++, 0 )

int vec_insert_(char **data, int *length, int *capacity, int memsz,
                 int idx) {
  int err = vec_expand_(data, length, capacity, memsz);
  if (err) return err;
  memmove(*data + (idx + 1) * memsz,
          *data + idx * memsz,
          (*length - idx) * memsz);
  return 0;
}
\end{lstlisting}
\end{minipage}
\caption{Implementation of the {\tt vec\_insert} function that adds the {\tt val} value in the {\tt idx} index of the {\tt vec} structure. We omit the function \texttt{vec\_expand\_} for simplicity: it reallocates the vector if it needs to be expanded.}
\label{fig:ansi-c-example}
\end{figure}

%----------------------------------------------------------------
\subsection{Checking for Relevant Code Changes}
\label{sec:equivalence}
%----------------------------------------------------------------

We begin from the principle that if a modified version of a component is computationally equivalent to its older version, then it is not necessary to prove that all properties that hold for the old version still hold for the modified one. Thus, we use equivalence checking to check whether the modified components need to be re-verified. Naturally, proving the equivalence of two functions is in general undecidable~\cite{Strichman09}, and the effort we spend in trying to do so might be wasted. However, such an approach can potentially reduce the immediate verification effort, since proving the equivalence of two function versions can be less expensive than re-verifying the function~\cite{Strichman09}. In addition, by proving that two versions of a function are computationally equivalent, we eliminate the effort to re-verify any other function that depends on it (unless that function has been changed as well). Therefore, this approach limits the propagation of changes through the system and, consequently, reduces the effort to overall system verification.

The equivalence check will happen in two steps: a (1) fast and imprecise abstract syntax tree (AST) structural equivalence check~\cite{Ramos:2011}, and a (2) slow and precise formal check e.g. bounded model checking (BMC). In the AST structural equivalence check, easy cases will be caught without the need to formally verify it, e.g., a function is renamed and the call sites are updated, or comments are added to a function body. If the AST is structurally not equivalent, we then encode the old and the new functions, and check if they are equivalent for the same inputs. A time limit is set for the formal check since it is more useful to spend time running the regression tests than checking their equivalence; if the time limit is exhausted, we assume they are not equivalent and start the tests.

In our illustrative project \texttt{vec} we find commits that would benefit from our approach. In commit $40d5cc17$\footnote{https://github.com/rxi/vec/commit/40d5cc17ea41923c66286078bae82cc09c6458f7}, the developer changes that name of a macro \texttt{vec\_absindex} used in an early version of the function shown in Fig.~\ref{fig:ansi-c-example}, and in commit $7d8588bc$\footnote{https://github.com/rxi/vec/commit/7d8588bc96c4c7aa68beb38f15704bd6135c0a5e}, the developer removes the support for negative indexes when accessing arrays. In the former, the ASTs would found to be equivalent, neither triggering the next formal check nor starting the tests, while in the latter, the formulas would be found to be not equivalent by the formal check, triggering the regression tests.

{\bf Open Challenges.} There are many techniques that could be applied to perform equivalence checking such as SYMDIFF~\cite{symdiff} and CORK~\cite{Lopes2016} tools or through directed incremental symbolic execution (DiSE)~\cite{Person:2011}; in future, we will evaluate their performance in this CFV setting. We will also exploit this module by generating test cases from code changes~\cite{Godefroid:2011}.

\subsection{Model Checking Generalized Tests}
\label{sec:generalizing}

%After detecting new and/or modified components, we use the existing regression test cases to reduce the state space to be explored by the model checker. In that sense,
It is of paramount importance a software project follows two key best-practice principles: {\it (i)} keep the project as modular as possible and create short functions that focus on one particular objective {\it (ii)} provide at least one regression/unit test for every function. Such an approach is key to a successful compositional analysis of the software project, where the combination of the analysis result of its parts represents the analysis result of the whole.

After pruning the unmodified components, we only focus on the existing regression test cases related to the modified ones, in order to reduce the state space to be explored by the model checker. However, we do not generate new concrete values for the test cases with the purpose of maximizing the code coverage. Instead, we combine existing test cases with non-deterministic input values to maximize the coverage of this verification. The use of regression tests also help to reduce the state space by breaking the global model (containing the entire program) into local models (containing only the functions under verification) and generate on-demand the reachable states to be visited by the model checker, starting with the state described by the test case. This reduces the number of paths and variables to be considered during model checking.

In our illustrative project \texttt{vec}, by measuring the number of linearly independent paths in all functions, i.e., the project's cyclomatic complexity~\cite{Bang:2015}, we clearly see the benefit of focusing on the regression tests. In the case of {\tt vec}, the entire system has a cyclomatic complexity of 24; in contrast, its regression tests have an average cyclomatic complexity of 1. %During dynamic verification, we are not able to find any bug in the circular buffer implementation with these. However, the implementation is flawed: the array \emph{buffer} is declared to be of type \emph{char}[] (see line~1 in Figure~\ref{fig:continuous-integration}) but we assign an element \emph{b} of type \emph{int} (see line~14). The test cases do not uncover this error because they happen to use only integer values that can safely be cast to a \emph{char}.
Through BMC, we can check for all possible paths in the implementation shown in Fig.~\ref{fig:ansi-c-example}, by non-deterministically assigning a value for each function parameter ({\it i.e.}, {\tt pos}, and {\tt val}) assuming a valid initialized structure ({\it i.e.}, {\tt v}).
%However, this approach can lead to false negatives as the non-deterministic choice of values for program variables may force the exploration of paths that are infeasible in the original program.
Rather than modifying the program, we modify the regression tests and replace the concrete input values by non-deterministic choices. Here, we replace the series of function invocations with a non-deterministic one (see lines~5--7 of Fig.~\ref{figure:generalization}). We can try to get full coverage in this particular module because we already pruned the state space by only selecting the modified parts of the system.%In order to block parts of the search space, we use the given concrete values from all regression tests and combine the respective values into a single interval for each variable or array element (see Fig.~\ref{figure:generalization}); here we assume that all ``obvious'' boundary values are used in some of the tests (\textit{e.g.}, using boundary-value analysis), so that we force the model checker towards the ``unobvious'' errors. %In the example, we thus add an  \emph{assume}-statement such as \emph{assume}(\emph{ssr}[0]$<$1 $\|$ \emph{ssr}[0]$>$43) as shown in Figure~\ref{figure:unit-test-buffer-with-assume} and we are then able to find two bugs related to overflow and underflow. Table~\ref{table:combine-the-concrete-values-into-a-sinle-interval} shows all concrete values to check dynamically the circular buffer and that we used to derive the single intervals (shown in Figure~\ref{figure:unit-test-buffer-with-assume}).

\begin{figure}[!h]
\centering
\begin{subfigure}{.49\textwidth}
\begin{lstlisting}[escapechar=^]
test_section("vec_insert");
vec_int_t v;
vec_init(&v);
int i;
for (i = 0; i < 1000; i++)
  vec_insert(&v, 0, i);
test_assert(v.data[0] == 999);
test_assert(
  v.data[v.length - 1] == 0);
vec_insert(&v, 10, 123);
test_assert(v.data[10] == 123);
test_assert(v.length == 1001);
vec_insert(&v, v.length - 2, 678);
test_assert(v.data[999] == 678);
test_assert(
  vec_insert(&v, 10, 123) == 0);
vec_insert(&v, v.length, 789);
test_assert(
  v.data[v.length - 1] == 789);
vec_deinit(&v);
\end{lstlisting}
\caption{Original test.}
\label{figure:regression-test}
\end{subfigure}
\hfill
\begin{subfigure}{.44\textwidth}
\begin{lstlisting}[escapechar=^]
test_section("vec_insert");
vec_int_t v;
vec_init(&v);
int val = nondet_int();
size_t pos = nondet_size_t();
vec_insert(&v, pos, val);
test_assert(v.data[pos] == val);
vec_deinit(&v);
\end{lstlisting}
\caption{Generalized version.}
\label{figure:generalization}
\end{subfigure}
\caption{Generalization of the regression test for the function shown in Fig.~\ref{fig:ansi-c-example}.}
\label{figure:kind-examples}
\end{figure}
{\bf Open Challenges.} Our main difficulty here is how to deal with false negatives as the non-deterministic choice of values for program variables may force the exploration of paths that are infeasible in the original program. So, we need to find a balance between coverage and soundness. We also need to increase automation as much as possible. One may combine techniques to automatically generate tests based on counterexamples~\cite{Dirk:2018} or source code~\cite{Christakis:2017}. We will also increase the power of this analysis by using conditional verifiers~\cite{Beyer:2018} or applying different model checking approaches ({\it i.e.}, explicit-state).

%-----------------------------------------
\section{Related Work}
\label{sec:related-work}
%----------------------------------------

Fitzgerald and Stol~\cite{FITZGERALD2017176} present a holistic overview of the activities related to continuous software engineering, which includes continuous testing and verification. Although they do not propose a new approach, they highlight the importance of continuous (and automatic) testing and verification in the context of DevOps. Interestingly, Beyer and Lemberger~\cite{dirk2017} perform a comparison between software testers and software model checkers, which shows that model checkers are mature enough to be used in practice (they even outperform testing tools), and the combination of both techniques could lead to even better results. Indeed, there are many reports of successful attempts that use formal techniques in large software systems.

For instance, Klein {\it et al.}~\cite{Klein:2018} show how to scale formal proofs based on software architecture to real systems at low cost; % Holzmann present Cobra~\cite{Holzmann2017}, a static code analyzer that can effectively handle large code bases;
Godefroid, Levin, and Molnar~\cite{Godefroid:2012} describe the remarkable impact of SAGE tool, which performs dynamic symbolic execution to hunt for security issues in Microsoft applications; Cordeiro, Fischer, and Marques-Silva~\cite{CordeiroFM10} as well as Yin and Knight~\cite{NFM2010:YiKn} propose approaches to conduct formal verification of large software systems. Furthermore, there are two important studies that tackle the combination of formal techniques with continuous integration, which led to promising results and reflect the need and scientific challenges in the industry to follow this road. First, Chudnov {\it et al.}~\cite{Chudnov2018} describe how Amazon Web Services (AWS) prove the correctness of their Transport Layer Security (TLS) protocol implementation, and how they use CI tools to keep proving the software properties during its lifetime. Similarly, O'Hearn~\cite{OHearn:2018} presents Infer, a static analyzer used at Facebook following a continuous reasoning approach. Neither Chudnov~{\it et al.} nor O'Hearn try to handle model checking in a continuous process; the latter states this as an open challenge for the community.

These cases highlight the impact of formal techniques in real software systems; however, they do not present guidelines to generalize these approaches to a wide range of software projects, which could lead to a significant adoption of formal techniques by practitioners. Thus, there is still an open-call for approaches that could potentially popularize formal techniques in software engineering practices.

\section{Conclusions and Future Work}
\label{sec:conclusion}
%--------------------------------------------------------------

Model checking of entire systems is usually not feasible for many industrial applications due to the state-space explosion problem, however, one of the scalability challenges can be solved through leveraging changes to the system. Thus, we propose CFV, an approach with the potential to detect software vulnerabilities by combining dynamic and static verification to reduce the state space. This potential propels us to further research this topic: we are currently developing an automated software tool to tackle the
%\ As a result, the continuous formal verification approach and the combination of different encodings and solvers allowed us to explore more exhaustively the state space of the program.\
%Controlled e
%Experiments using the NXP set-top box software with more than $10$K of LOC shows that this approach can potentially improve the error-detection capability and reduce the verification time.
key challenges of equivalence checking and test case generalization, so it can be applied to large open-source projects. We are also working in close collaboration with software developers at Amazon Web Service and Samsung with the goal of integrating our automated reasoning tool into their workflow, thus increasing adoption of formal methods in industry.
%However, the advantage of the continuous verification approach is not as substantial as expected. If there would be more functional correctness assertions in our case study, re-verification would be more expensive and the continuous verification approach would then be more advantageous.

\bibliographystyle{splncs}
\bibliography{references}

\end{document}